\documentclass[prb,aps, twocolumn, amsmath,amssymb,superscriptaddress]{revtex4}
\usepackage{graphicx} % Required for inserting images
\usepackage{float}
\usepackage{amsmath}
\usepackage{amssymb}
\usepackage{amsfonts}
\usepackage{euscript}
\usepackage{enumerate}
\usepackage{hhline}
\usepackage{pslatex}
\usepackage{tabularx}
\usepackage{braket}
\usepackage[usenames,dvipsnames]{xcolor}

\usepackage[colorlinks,linkcolor=red,anchorcolor=black,citecolor=black]{hyperref}
\usepackage{soul}
\usepackage{tabularx}
\usepackage{array}
\usepackage{booktabs}
\usepackage{xcolor}

\makeatletter
\renewcommand{\@biblabel}[1]{#1. }
\renewcommand{\@dotsep}{500}
\renewcommand{\@pnumwidth}{0em}
\renewcommand{\l@figure}[2]{% #1 is e.g. Figure 1 + caption, #2 is pg.
\newcommand{\ysnote}[1]{{\color{red} #1}}
\@dottedtocline{1}{1.5em}{2em}{Figure #1}{}\vspace{15pt}}

\begin{document}

\title{Structured Cavity Quantum Electrodynamics}

\author{Shunfa Liu}
\thanks{These authors contributed equally to this work.}
\affiliation{State Key Laboratory of Optoelectronic Materials and Technologies, School of Physics, Sun Yat-sen University, Guangzhou 510275, China.}

\author{Jiantao Ma}
\thanks{These authors contributed equally to this work.}
\affiliation{State Key Laboratory of Optoelectronic Materials and Technologies, School of Physics, Sun Yat-sen University, Guangzhou 510275, China.}

\author{Hanqing Liu}
\thanks{These authors contributed equally to this work.}
\affiliation{State Key Laboratory of Optoelectronic Materials and Devices, Institute of Semiconductors, Chinese Academy of Sciences, Beijing 100083, China.}
\affiliation{Center of Materials Science and Optoelectronics Engineering, University of Chinese Academy of Sciences, Beijing 100049, China.}

\author{Yangpeng Wang}
\affiliation{State Key Laboratory of Optoelectronic Materials and Technologies, School of Physics, Sun Yat-sen University, Guangzhou 510275, China.}

\author{Xueshi Li}
\affiliation{State Key Laboratory of Optoelectronic Materials and Technologies, School of Physics, Sun Yat-sen University, Guangzhou 510275, China.}

\author{Haiqiao Ni}
\affiliation{State Key Laboratory of Optoelectronic Materials and Devices, Institute of Semiconductors, Chinese Academy of Sciences, Beijing 100083, China.}
\affiliation{Center of Materials Science and Optoelectronics Engineering, University of Chinese Academy of Sciences, Beijing 100049, China.}

\author{Zhichuan Niu}
\affiliation{State Key Laboratory of Optoelectronic Materials and Devices, Institute of Semiconductors, Chinese Academy of Sciences, Beijing 100083, China.}
\affiliation{Center of Materials Science and Optoelectronics Engineering, University of Chinese Academy of Sciences, Beijing 100049, China.}

\author{Kai Zou}
\affiliation{School of Precision Instrument and Optoelectronic Engineering, Key Laboratory of Optoelectronic Information Science and Technology, Ministry of Education, Tianjin, China.}

\author{Yun Meng}
\affiliation{School of Precision Instrument and Optoelectronic Engineering, Key Laboratory of Optoelectronic Information Science and Technology, Ministry of Education, Tianjin, China.}

\author{{Xiaolong Hu}}
\affiliation{School of Precision Instrument and Optoelectronic Engineering, Key Laboratory of Optoelectronic Information Science and Technology, Ministry of Education, Tianjin, China.}
\affiliation{College of Information Science and Electronic Engineering, Zhejiang University, Hangzhou 310027, China}

\author{Xuehua Wang}
\thanks{wangxueh@mail.sysu.edu.cn}
\affiliation{State Key Laboratory of Optoelectronic Materials and Technologies, School of Physics, Sun Yat-sen University, Guangzhou 510275, China.}
\affiliation{Quantum Science Center of Guangdong-Hong Kong-Macao Greater Bay Area, Shenzhen 518045, China.}

\author{Jin Liu}
\thanks{liujin23@mail.sysu.edu.cn}
\affiliation{State Key Laboratory of Optoelectronic Materials and Technologies, School of Physics, Sun Yat-sen University, Guangzhou 510275, China.}
\affiliation{Quantum Science Center of Guangdong-Hong Kong-Macao Greater Bay Area, Shenzhen 518045, China.}
\date{\today}

\begin{abstract}
\noindent \textbf{A cavity quantum electrodynamics (cQED) system consisting of a confined single photon and a single quantum emitter serves as a fundamental block for quantum optics and photonic quantum technologies. The canonical optical mode employed in the conventional cavity quantum electrodynamics features a uniform polarization distribution, leading to the scalar light-matter interaction in most existing experiments. Despite the rapid progress in the generation of structured light with spatially varied polarizations, the structured light-matter interaction, especially at the single quanta level, is highly intriguing yet largely unexplored. Here, we present the structured light-matter interaction at the single-photon level in a semiconductor cavity quantum electrodynamics system. Four distinct structured cavity modes that are spectrally close to each other are constructed in a micropillar cavity. By spatially locating a single epitaxial quantum dot (QD) at the periphery of a semiconductor micropillar cavity and spectrally tuning the QD emission wavelength into the resonances of the structured cavity modes, cavity-enhanced single-photon emissions with spin-locked chiral orbital angular momentum (OAM) and engineerable spin-orbit entanglements are achieved within a single wavelength-scale device. Our work opens an unexplored paradigm of structured quantum light-matter interactions and may further advance chiral quantum optics and high-dimensional photonic quantum technology.} 
\end{abstract}

\maketitle

\begin{figure*}
	\includegraphics[width=1\linewidth]{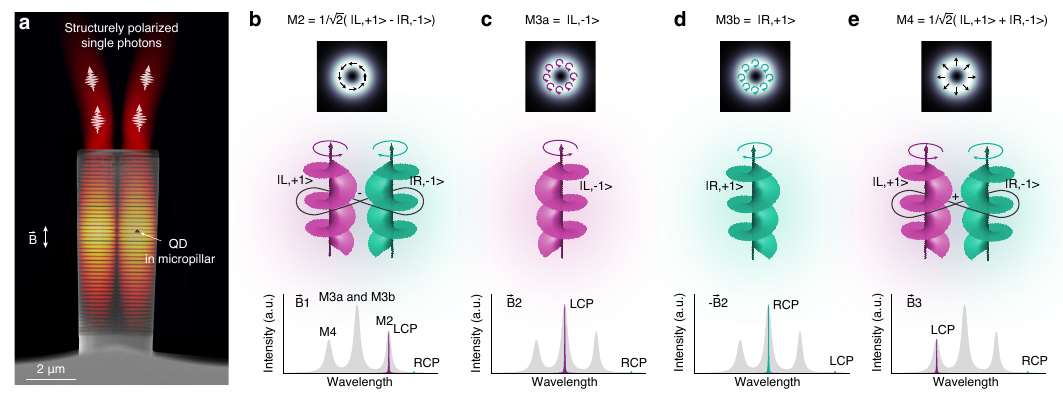}
	\caption{\textbf{Structured quantum light-matter interaction in a semiconductor cQED system.} (a) Schematics of the employed cQED system in which a single QD is deterministically coupled to different structured cavity modes of the micropillar. (b) Spin-orbit entanglement in the form of M2 = $\frac{\sqrt{2}}{2} ( \vert \rm{L},+1\rangle - \vert \rm{R},-1\rangle) $ by tuning the QD into the resonance of M2. (c) Chiral OAM single photon-emission (left-hand circular polarization with OAM of -1) by tuning QD into the resonance of M3a. (d) Chiral OAM single photon-emission (right-hand circular polarization with OAM of 1) by tuning QD into the resonance of M3b. (e) Spin-orbit entanglement in the form of M4 = $\frac{\sqrt{2}}{2} ( \vert \rm{L},+1\rangle + \vert \rm{R},-1\rangle) $ by tuning the QD into the resonance of M4.}
	\label{fig:Fig1}
\end{figure*}

The interaction between a single photon and a single quantum emitter is the elementary process~\cite{dutra2005cavity} in quantum physics~\cite{loudon2000quantum} and photonic quantum technology~\cite{o2009photonic,wang2020integrated}. In an integrated cQED system, a single photon is confined in a wavelength-scale cavity for a long time, resulting in strongly enhanced interactions with a single quantum emitter. Modern chip-scale cQED platforms have been successfully employed to build high-performance quantum light sources in the weak coupling regime~\cite{somaschi2016near,he2017deterministic,wang2019towards,liu2019solid,wang2019demand,uppu2020scalable,tomm2021bright} and construct quantum logic gates in the strong coupling regime~\cite{reinhard2012strongly,volz2012ultrafast,sun2018single}, serving as a central element in photonic quantum technology covering quantum communication~\cite{gisin2007quantum}, quantum computing~\cite{o2007optical,zhong2020quantum,madsen2022quantum} and quantum metrology~\cite{giovannetti2011advances}. The standard cavity modes employed in cQED exhibit spatially uniform polarizations, leading to scalar light-matter interactions in most of the cQED experiments to date ~\cite{somaschi2016near,he2017deterministic,wang2019towards,uppu2020scalable,tomm2021bright}. The last decades have witnessed the burgeoning of structured light whose polarization, amplitude, phase and even OAM can be tailored on-demand, enabling a series of breakthroughs in optical communication, quantum information processing, optical manipulation and super-resolution imaging~\cite{forbes2021structured,he2022towards}. Despite extensive efforts in the creation of different types of structured light~\cite{wang2021generation,wang2020vectorial,carlon2019optically}, the investigation of the interactions between structured light and matter is highly intriguing yet largely unexplored, especially at the single quanta level. Along this exciting direction, we have witnessed the continuous development of OAM single-photon sources based on solid-state quantum emitters\cite{chen2021bright,Ma2022OnChip,Wu2022Room,Liu2023OnChip,zhao2023high,Liu2024Ultracompact,Ma2026HighPerformance}, see more details in Extended Data Fig. E1, yet all the existing devices require extra optical elements to introduce effective phase modulations. Another prominent example of structured light-matter interaction at the single-photon level is the emerging chiral quantum optics in which local polarization states of structured light lock to its propagation direction, leading to propagation-direction-dependent quantum emission~\cite{lodahl2017chiral,suarez2025chiral}. Beyond emission chirality, the multiple degrees of freedom associated with the structured light could also be employed to produce non-classical radiation with on-demand OAM and engineerable spin-orbit couplings by using quantum meta-optics devices~\cite{stav2018quantum}.

In this work, we explore the structured light-matter interaction in the cQED regime. By leveraging the optical spin-orbit coupling effect~\cite{dufferwiel2015spin,sala2015spin,carlon2019optically,whittaker2019effect,liu2021dual} in a high-quality semiconductor micropillar cavity, we successfully realize a set of highly confined optical modes whose spatial profile, phase and polarization feature spatial inhomogeneities. A single epitaxial QD is deterministically placed in the specific location of the structured cavity mode, enabling the vectorial light-matter coupling in the semiconductor cQED system. By spectrally tuning the QD into the resonances of structured cavity modes under magnetic fields, single-photon emission with a $\rm g^{(2)}(0) = 0.062(1)$ and a Purcell factor of $\sim$4 are obtained, exhibiting chiral OAM with a mode purity as high as 0.902 and engineerable spin-photon entanglement with a fidelity up to 0.952. Our work brings the structured light-matter interaction into the previously unexplored cQED regime and may directly boost the development in advanced quantum technology, such as high-dimensional quantum communications\cite{cozzolino2019high} and high-density quantum memories~\cite{dong2023highly}.

\section{Principles}

The schematics of our device are presented in Fig.~\ref{fig:Fig1}(a). An InAs QD is intentionally placed off from the centre of a high-quality micropillar cavity consisting of 30(20) pairs of bottom(top) $\rm GaAs/Al_{0.9}Ga_{0.1}As$ distributed Bragg reflector (DBR) and a $\lambda$-cavity. The QD, operating as a high-performance single-photon emitter, is coupled to the confined cavity modes with spatially varied polarizations, as schematically shown in Fig~\ref{fig:Fig1}(a). The micropillar cavity supports four confined modes exhibiting characteristics of structured light. The single photons emitted from the QD are efficiently funneled to different structured cavity modes via the Purcell effect, resulting in the emission of single photons with spin-locked chiral OAM (Fig.~\ref{fig:Fig1}(c,d)) or engineerable spin-orbit entanglements (Fig.~\ref{fig:Fig1}(b,e)), depending on which cavity mode the emitter is coupled to.

\begin{figure*}
	\includegraphics[width=1\linewidth]{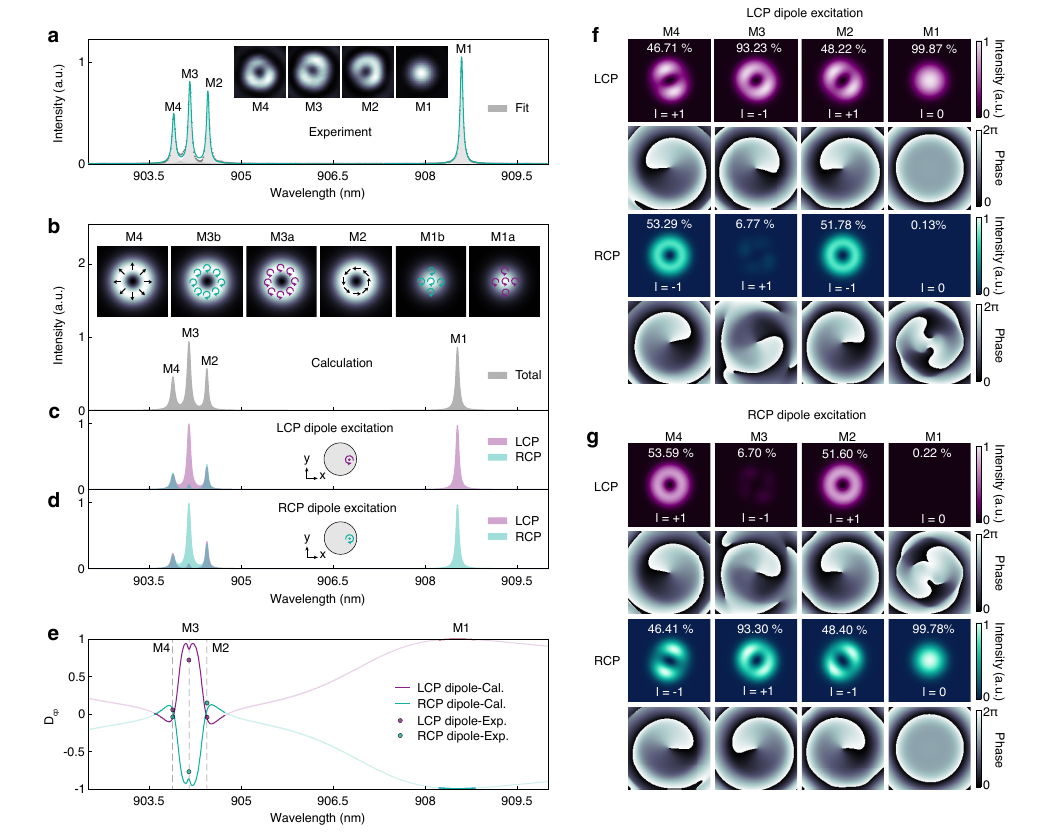}
	\caption{\textbf{Structured cavity modes and their polarization-selective excitations.} (a) Measured PL spectrum of the micropillar cavity with an embedded QD under a high-excitation power. Inset: intensity profile of each cavity mode. (b) Calculated intensity and polarization profiles (the first row) of the cavity modes and the simulated spectrum (the second row) of the micropillar. The selective excitation of the cavity modes by using an LCP (c) and RCP (d) dipole source. (e) Calculated (solid lines) and measured (dots) $\rm{D_{cp}}$ for the different cavity modes. (f,g) Calculated  intensity profiles and phase distributions of the near-field cavity modes excited by an LCP and an RCP dipole source, respectively, from which the $\rm{D_{cp}}$ for M1-M4 in (e) can be extracted.}
	\label{fig:Fig2}
\end{figure*}

\section{Characterization of structured cavity modes}

In Fig.~\ref{fig:Fig2}(a), we present the measured photoluminescence (PL) spectrum of the micropillar cavity obtained under a high excitation power, in which the fundamental mode (denoted by M1) and higher-order modes (denoted by M2-M4) of the cavity can be clearly identified (see Extended Data Fig. E2(b)). The intensity profile of M1 is a Gaussian distribution, indicating a uniform polarization, while M(2-4) features a doughnut intensity profile, suggesting the creation of structured optical modes. The simulated full spectrum of the micropillar cavity is shown in Fig.~\ref{fig:Fig2}(b), which is in excellent agreement with the experiment of Fig.~\ref{fig:Fig2}(a). From the simulations, we can further reveal that the M1 and M3 are polarization-degenerate while the M2 and M4 are azimuthally and axially polarized, respectively, as shown by the insets of Fig.~\ref{fig:Fig2}(b). In our system, one of the degenerate modes, M1a/M3a with left-hand circular polarization (LCP) or M1b/M3b with right-hand circular polarization (RCP),  can be selectively excited by placing a LCP or RCP dipole source in the cavity, as shown in Fig.~\ref{fig:Fig2}(c.d). The mode degeneracy of M3 are presented in Extended Data Fig. E2(b). We further calculate the degree-of-circular-polarization $\rm{D_{cp}}$ (see Method) of each cavity mode in Fig.~\ref{fig:Fig2}(e), in which M1 and M3 exhibit very high $\rm{D_{cp}}$ values when excited by circularly polarized dipole sources. Such mode selection processes are clearly visualized in Fig.~\ref{fig:Fig2}(f,g). The M1a(b) and M3a(b)  can only be excited by the embedded LCP(RCP) dipole source, while the M2 and M4 show no appreciable mode selectivity.

\begin{figure*}
	\begin{center}
		\includegraphics[width=1.0\linewidth]{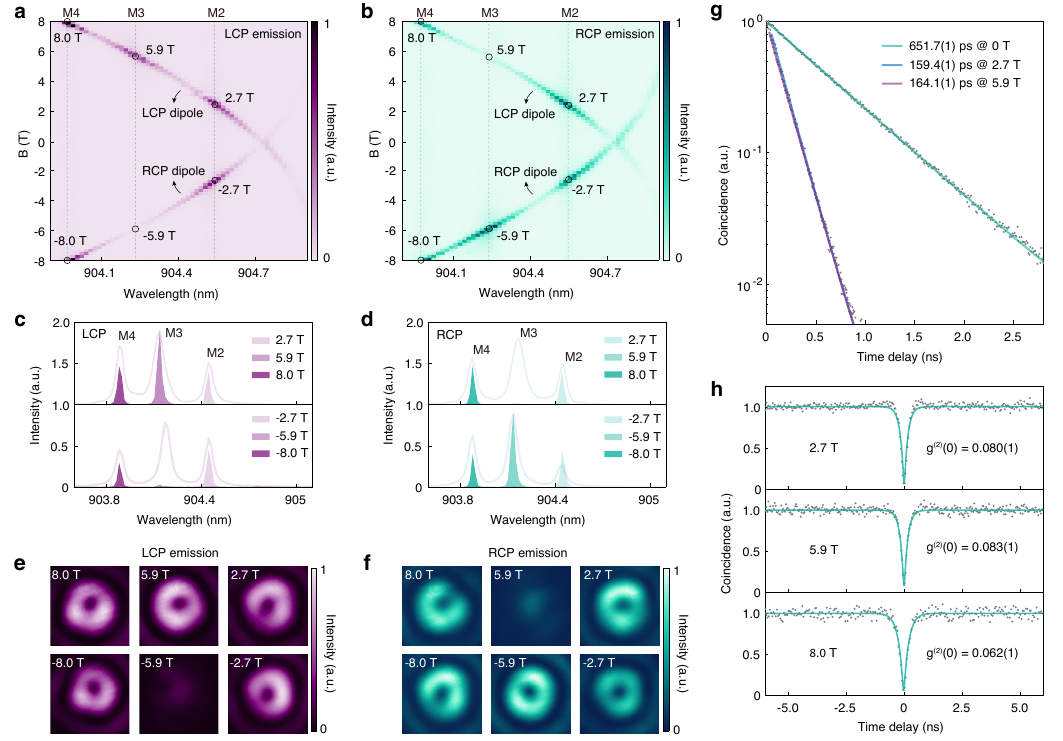}
		\caption{\textbf{Experimental verification of structured light-matter coupling at the single-photon level.} (a) LCP and (b) RCP component of the magnetic field-dependent PL spectra for QD under low-power pulsed excitation. The vertical dash lines denote the cavity resonance of M2, M3 and M4. (c) LCP component of the LCP dipole emission at 2.7 T, 5.9 T, 8.0 T and -2.7 T, -5.9 T, -8.0 T. (d) RCP component of the LCP dipole emission at 2.7 T, 5.9 T, 8.0 T and -2.7 T, -5.9 T, -8.0 T. (e) LCP component of near-field profiles for the LCP dipole emission at 2.7 T, 5.9 T, 8.0 T and -2.7 T, -5.9 T, -8.0 T. (f) RCP component of near-field profiles for the LCP dipole emission at 2.7 T, 5.9 T, 8.0 T and -2.7 T, -5.9 T, -8.0 T. (g) Lifetimes of the LCP dipole measured at 0 T, 2.7 T and 5.9 T, showing pronounced enhancement of spontaneous emission rate by a factor of $\sim$4. (h) Second-order correlations of the LCP dipole emission measured 2.7 T, 5.9 T and 8.0 T, showing highly pure single-photon characteristics.}
		\label{fig:Fig3}
	\end{center}
\end{figure*}

\section{Structured quantum light-matter coupling}
We now experimentally couple a single QD to the structured cavity modes and demonstrate the mode selection by tuning the QD wavelength to different cavity resonances. Under the magnetic field in the Faraday configuration, the optically-excited neutral exciton states in the QD is split into an LCP and an RCP dipole source and their wavelengths can be continuously tuned by varying the strength of the magnetic field~\cite{barik2018topological}, see the polarization-resolved spectra of a QD in a planar cavity in Extended Data Fig. E2(c-e). We note that although the same Zeeman tuning technique has been used to create quantum photonic skrymions\cite{Ma2025Nanophotonic}, both scientific motivation and technological implementation here are appreciable distinct from our previous work, see Extended Data Tab. E1 The polarization-resolved PL spectra of the single QD under different magnetic fields are presented in Fig.~\ref{fig:Fig3}(a,b), in which three appreciable emissions can be identified when the LCP dipole source is resonant to the cavity modes at 2.7 T, 5.9 T and 8.0 T. Ideally, the LCP dipole does not emit RCP photons and vice versa. However, in the LCP spectra we observe two pronounced RCP emissions from the LCP dipole at -2.7 T and -8 T in Fig.~\ref{fig:Fig3}(a,c). These unexpected emissions arise from efficient coupling of the RCP dipole to M2 and M4 with vectorial polarizations, leading to coherent conversion from RCP to LCP emission. In contrast, there is negligible LCP emission when the RCP dipole is coupled to M3. Similar behaviors are observed in the RCP spectra, in which LCP emission from the RCP dipole is observed at 2.7 T and 8.0 T while no RCP emission from the LCP dipole is observed at 5.9 T, as shown in Fig.~\ref{fig:Fig3}(b,d). The measured cavity emissions feature different $\rm{D_{cp}}$, as shown by the points in Fig.\ref{fig:Fig2}(e), which is in good agreement with the calculation.  We further present the polarization-resolved near-field profiles of the dipoles at different magnetic fields, as shown in Fig.~\ref{fig:Fig3}(e,f). When coupled to the structured cavity modes, the emission profiles of QD become a doughnut-shaped beam and the emission exhibits strong polarization dependence at the resonance of M3. In addition, the efficient couplings between the QD and the cavity modes are revealed from the time-resolved measurements, featuring Purcell factors of $\sim$4, as shown in Fig.~\ref{fig:Fig3}(g). The measured Purcell factor here is lower than the simulated value, due to the fact that the QD's spatial position is not ideal and its emission wavelength is still not fully detuned from the cavity mode at the highest magnetic field. We have obtained a higher Purcell factor up to 12.53 in a different device, see Extended Data Fig. E3 and Extended Data Tab. E2. Finally, all the cavity-enhanced emissions measured at the saturation powers are verified as highly pure single photons by the Hanbury-Brown-Twiss interference measurements, as presented in Fig.~\ref{fig:Fig3}(h). The chiral OAM emission is very robust to the position variation of QD (see Extended Data Fig. E4(a-c)), therefore such an effect has been clearly observed in different devices in our sample (see Extended Data Fig. E3). Regarding the source brightness, the extraction efficiencies for M2-M4 are very close to that of M1, which is widely used for bright single-photon sources, see Extended Data Fig. E4(d).  Therefore, the structured light-matter interactions at the single-photon level are clearly demonstrated from the polarization-dependent PL measurements under varied magnetic fields. We note that the structured light-matter interaction in this work could also be exploited in other types of microcavities with careful photonic engineering, such as open cavities\cite{dufferwiel2015spin} and photonic crystal cavities\cite{Chang2017Cylindrical,Gao2020Vector,Fong2021Chiral}, by locating QDs at the anti-nodes of the structured cavity modes.

 \begin{center}
	\begin{figure*}
		\begin{center}
			\includegraphics[width=\linewidth]{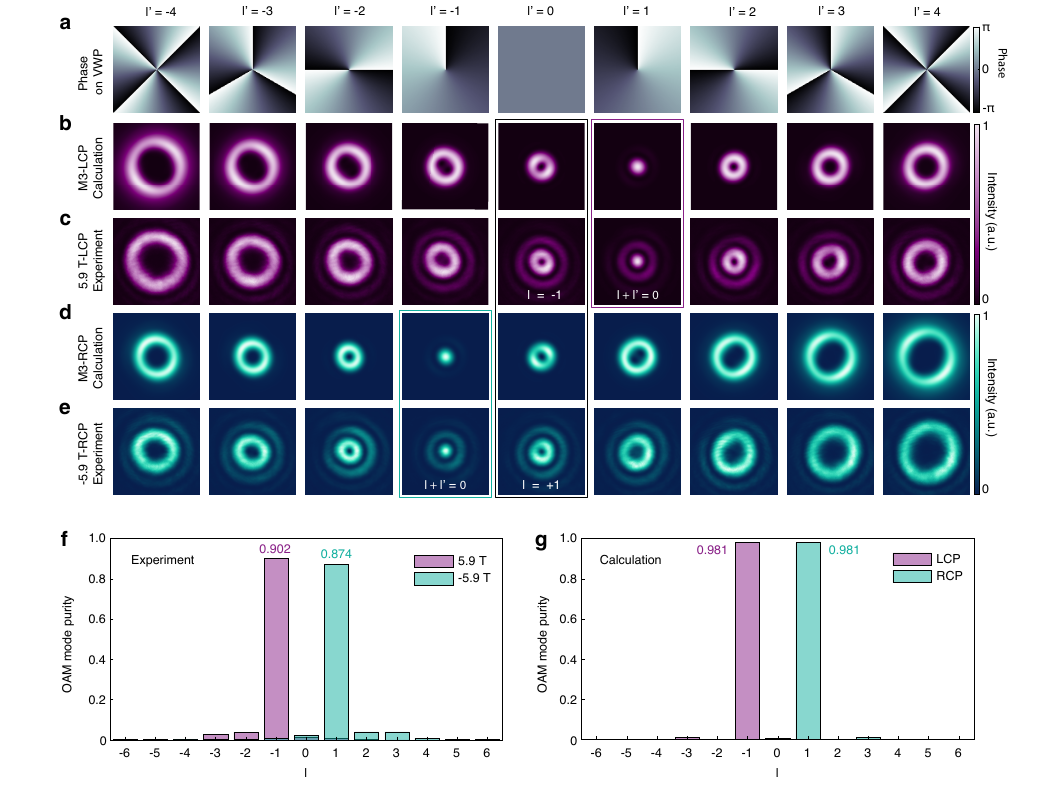}
			\caption{\textbf{Single-photon emission with spin-locked chiral OAMs.} (a) phase profiles carried by the VWPs. calculated (b) and measured (c) near-field normalized intensity distributions of the LCP dipole emitter coupled to M3a. calculated (d) and measured (e) near-field normalized intensity distributions of the RCP dipole emitter coupled to M3b. Calculated (a) and measured (b) mode purity of the spin-locked chiral OAMs.The unexpected outside ring patterns in the experiment are due to the fact that the objective is slightly off from the focus in the measurement.}
			\label{fig:Fig4}
		\end{center}
	\end{figure*}
\end{center}

\section{Spin-lock chiral OAM single-photon emission}
M3a and M3b are confined optical modes featuring spin-locked chiral OAM. When coupled to circularly polarized dipole sources, the resulting single photons carry spin-dependent OAM. We analyze the near-field emission patterns of the emitted single photons by projecting them onto a vortex wave plate (VWP) serial with variable topological charge $l'$~\cite{cai2012integrated,chen2021bright,sroor2020high}, as shown in Fig.~\ref{fig:Fig4}(a). At 5.9 T, the LCP dipole emitter is resonant with M3a, and the single-photon emission carries an OAM with a topological charge of -1, as shown in Fig.~\ref{fig:Fig4}(b,c). When reversing the magnetic field to -5.9 T, the RCP dipole is coupled to the M3b, which gives rise to the single photons carrying OAM with a reversed topological charge of 1, as shown in Fig.~\ref{fig:Fig4}(d,e). OAM mode purities as high as 0.902 and 0.874 are extracted from the OAM spectrum as shown in Fig.~\ref{fig:Fig4}(f), which are very close to the theoretical value of 0.981 in Fig.~\ref{fig:Fig4}(g).  The simulated OAM purity is not reaching unity, probably due to the tiny cross-talk between the spectrally adjacent modes. It is worthwhile noting that single-photon vortices with large topological charges can also be explored by harnessing the higher-order modes of the micropillar cavity, see Extended Data Fig. E5. Demonstrating spin-locked chiral OAM single photons may immediately advance the emerging field of chiral quantum optics and high-dimensional quantum information processing. Furthermore, instead of using a slow magnetic field to switch the polarization of the QD dipole, it is feasible to employ phonon-assisted optical excitation to transfer the polarization states of the excitation laser to the emitted single photons from a charged exciton state~\cite{coste2023probing}, leading to all-optical chiral OAM quantum emission in an ultrafast manner.

 \begin{center}
	\begin{figure*}
		\begin{center}
 			\includegraphics[width=1\linewidth]{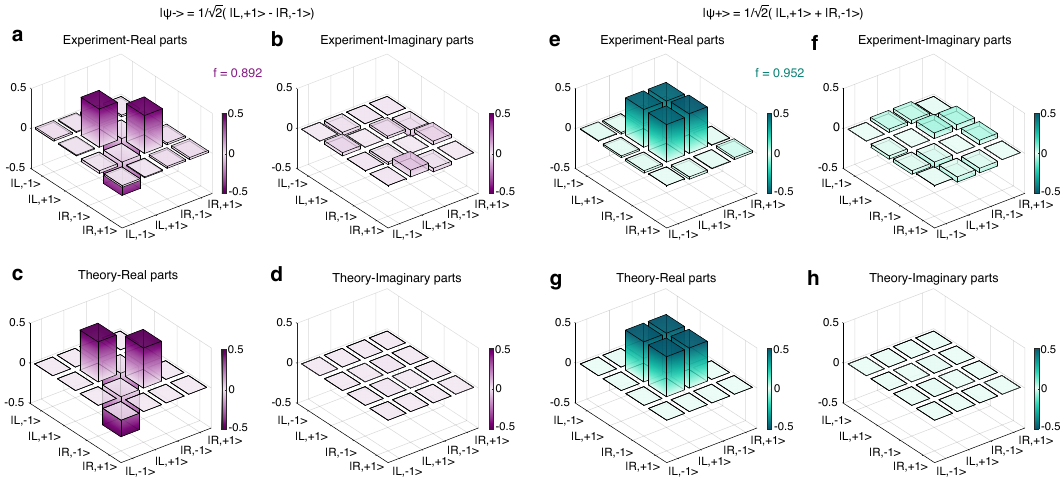}
			\caption{\textbf{Tunable spin-OAM entanglements in single photons.} Real (a) and imaginary (b) part of the experimentally reconstructed density matrix for the entanglement in the form of $\frac{\sqrt{2}}{2} ( \vert \rm{L},+1\rangle - \vert \rm{R},-1\rangle) $. (c) and (d) are the calculations for (a) and (b). Real (e) and imaginary (f) part of the experimentally reconstructed density matrix for the entanglement in the form of $\frac{\sqrt{2}}{2} ( \vert \rm{L},+1\rangle + \vert \rm{R},-1\rangle) $. (g) and (h) are the calculations for (e) and (f).}
			\label{fig:Fig5}
		\end{center}
	\end{figure*}
\end{center}

\section{Tunable spin-orbit entanglements via structured quantum light-matter interaction}
Finally, we investigate the generation of different types of spin-OAM entanglement in the emitted single photons. The inter-particle spin-OAM entanglement was previously demonstrated by propagating linearly-polarized single photons through bulk phase plates or ultrathin metasurfaces~\cite{stav2018quantum,suprano2023orbital}. However, the direct generation of such entanglement from a single chip-scale device remains extremely challenging due to the lack of an appropriate physical mechanism. We now solve this problem by exploring the structured quantum light-matter interaction in our semiconductor cQED system. When the LCP dipole emitter is resonant with M2, a maximal entangled Bell state, $\frac{\sqrt{2}}{2} ( \vert \rm{L},+1\rangle - \vert \rm{R},-1\rangle) $, is formed between the spin and OAM of the emitted single photons, as unequivocally demonstrated in the reconstructed density matrix from the quantum state tomography measurement (Fig.~\ref{fig:Fig5}(a,b)), which is in a very good agreement to the calculation in Fig~\ref{fig:Fig5}(c,d). By tuning the QD dipole to the resonance of M4, we successfully convert the Bell state to $\frac{\sqrt{2}}{2} ( \vert \rm{L},1\rangle + \vert \rm{R},-1\rangle) $, as shown in Fig.~\ref{fig:Fig5}(e-h). Our method is fundamentally different from the conventional approach of using linear optics elements, capable of dynamically and reversibly switching the entanglement between different Bell states. In addition, the other two Bell states can also be created by coupling linearly polarized dipole sources to M3a and M3b via phonon-assisted excitation on a charge exciton state~\cite{coste2023probing}. For the completeness of this work, we have further presented simulations for the couplings between linearly/circularly polarized dipoles and different cavity modes, as shown in Extended Data Fig. E6

\section{Conclusion}

To conclude, we have experimentally realized structured light-matter interactions at the single-photon level in a semiconductor cQED system. Four different types of high-Q structured and localized modes are constructed in a micropillar cavity. A single QD is deterministically placed at the periphery of the micropillar to ensure the spatial overlap between the QD and the structured light. Under magnetic fields, the single QD, serving as a circularly polarized dipole source, is tuned to the resonances of the high-order cavity modes to realize structured light-matter interaction, which is confirmed by both enhanced emission intensities and accelerated lifetimes of the exciton states. When coupled to the degenerate M3a and M3b, the emitted single photons from the QDs feature chiral OAM that locks to the spin of the circularly polarized dipole sources. We further realize the spin-OAM entangled Bell states by tuning the circularly polarized dipole source to the resonances of M2 and M4. Moving forward, it is highly desirable to pursue ultrafast all-optical single-photon emission carrying chiral OAM with an optically pumped charged exciton that is in resonance with M3a and M3b. The other two Bell states can also be created by using the charged exciton under phonon-assisted excitation when coupling to M2 and M4. Finally, we point out that the reversal process of chiral emission, i.e., chiral excitation, can be employed to transfer the photonic OAM of the excitation laser to the electron spin states in our structured CQED platform\cite{Fong2018Scheme}.  Our work provides a unique platform for investigating structured light-matter interactions at the single-quanta level, which is crucial for fundamental quantum science and may immediately boost the development of advanced quantum technologies, including chiral quantum optics and high-dimensional quantum information processing.

\vspace{0.8em}\noindent  \textbf{Data availability}

\noindent{The data that support the findings of this study are available within the paper and the Extended Data. Other relevant data are available from the corresponding authors on reasonable request.}

\vspace{0.8em}\noindent \textbf{Code availability}

\noindent{All codes produced during this research are available from the corresponding authors upon reasonable request.}

\vspace{0.8em}\noindent  \textbf{Acknowledgements}

\noindent{This research was supported by National Natural Science Foundation of China (12494600, 12494602, 62035017, 12361141824, 12574404, 12304409, 12474369); National Key Research and Development Program of China (2021YFA1400800); Natural Science Foundation of Guangdong Province (2023B1515120070, 2024B1515040013, 2026B1515020078); Guangdong Provincial Quantum Science Strategic Initiative (GDZX2206001, GDZX2306003, GDZX2503004); and the National Super-Computer Center in Guangzhou.}

\vspace{0.8em}\noindent  \textbf{Author Contributions}

\noindent J.~L. conceived the project; S.~F.~L., J.~L. and H.~Q.~L. designed the epitaxial structure and the devices; H.~Q.~L. and H.~Q.~N. grew the quantum dot wafers; S.~F.~L. and X.~S.~L. fabricated the devices; S.~F.~L., Y.~P.~W. and J.~T.~M. built the setup and performed the optical measurements; S.~F.~L. and  J.~T.~M. developed the model; K.~Z., Y.~M. and X.~L.~H. provided the SNSPD for lifetime and correlation measurement; S.~F.~L. and J.~L. analyzed the data; J.~L. and S.~F.~L. prepared the manuscript with inputs from all authors; J.~L., Z.~C.~N. and X.~H.~W. supervised the project.

\vspace{0.8em}\noindent  \textbf{Conflict of Interest}

\noindent The authors declare no competing interests.

\section{methods}
\noindent \textbf{Modeling:}
To calculate the cavity resonances, mode profiles, polarization and phase distributions of the micropillar, and to investigate the structured interaction between the high-order cavity modes and the quantum emitter, we modeled the structure using a three-dimensional finite-difference time-domain (3D-FDTD) method implemented in Ansys Lumerical FDTD. The diameter of the micropillar was set to \(2.6~\mu\mathrm{m}\), with 20/30 top/bottom distributed Bragg reflector (DBR) pairs. The circularly polarized quantum emitters were modeled by two orthogonal linearly polarized broadband dipole sources with a relative phase difference of \(\pm \pi/2\), corresponding to RCP and LCP dipoles, respectively. The quantum emitter was placed at the vertical center of the cavity layer and laterally displaced by \(650~\mathrm{nm}\) from the pillar center. The emitted field escaping from the cavity was recorded by a field monitor placed \(1~\mu\mathrm{m}\) above the top surface of the pillar. The radiation spectrum and properties of each mode are presented in Fig.~\ref{fig:Fig2}(b).  The spectrum presents four peaks between 903 nm and 909 nm, which correspond to the fundamental s-orbital mode(M1) and first-excited p-orbital modes(M2-M4) of the cavity with varied mode profile and polarization texture.

The degree of circular polarization, \(D_{\mathrm{cp}}\), was extracted from the simulated radiation spectrum recorded by the monitor. The electric field was decomposed into right- and left-circularly polarized components according to
$$
E_{\mathrm{RCP}} = \frac{1}{\sqrt{2}}(E_x - iE_y), \qquad
E_{\mathrm{LCP}} = \frac{1}{\sqrt{2}}(E_x + iE_y).
$$
The corresponding intensities, \(I_{\mathrm{RCP}}\) and \(I_{\mathrm{LCP}}\), were obtained by integrating the intensity of each circular polarization component around the target resonance. The degree of circular polarization was then calculated as
$$
D_{\mathrm{cp}} = \frac{I_{\mathrm{LCP}}-I_{\mathrm{RCP}}}{I_{\mathrm{LCP}}+I_{\mathrm{RCP}}}.
$$
This quantity characterizes the handedness selectivity of the emission enabled by the structured light-matter interaction. The properties of emission escaping from the cavity were extracted from the monitor at the top side. 

To understand the formation of the structured cavity modes in the micropillar, we start from an ideal weakly confined cylindrically symmetric cavity, where the paraxial approximation is valid. In this limit, the transverse eigenstates can be described by Laguerre--Gaussian-like modes in cylindrical coordinates $(r,\theta)$ together with circular polarization pseudospins, and the orbital and polarization degrees of freedom are approximately separable. Without spin--orbit coupling, the first excited $p$-like manifold with $|l|=1$ contains four degenerate basis states, $\vert \mathrm{R},+1\rangle$, $\vert \mathrm{R},-1\rangle$, $\vert \mathrm{L},+1\rangle$, and $\vert \mathrm{L},-1\rangle$. In a real etched micropillar, the finite lateral size and sidewall boundary conditions impose strong transverse confinement beyond the weak-confinement paraxial limit, which modifies the transverse mode profiles and breaks the separability between the orbital and polarization degrees of freedom. This lifts the ideal fourfold degeneracy and can be phenomenologically interpreted as a sidewall-confinement-induced spin--orbit coupling, or TE--TM-like splitting\cite{dufferwiel2015spin,sala2015spinorbit,carlon2019optically}.

Under this effective spin--orbit coupling, the first excited mode manifold is reorganized into three resonances, M2, M3, and M4. The lower- and higher-energy resonances correspond to two non-degenerate Bell-like vector vortex modes,
$$
\vert \mathrm{M2} \rangle =
\frac{1}{\sqrt{2}}
\left(
\vert \mathrm{L},+1\rangle
-
\vert \mathrm{R},-1\rangle
\right),
$$
and
$$
\vert \mathrm{M4} \rangle =
\frac{1}{\sqrt{2}}
\left(
\vert \mathrm{L},+1\rangle
+
\vert \mathrm{R},-1\rangle
\right),
$$
up to a global or convention-dependent phase. In contrast, the central M3 resonance forms a nearly energy-degenerate two-mode subspace. In the circular-polarization spin--OAM basis, the two M3 states are
$$
\vert \mathrm{M3a}_{\mathrm{L}} \rangle =
\vert \mathrm{L},-1\rangle,
$$
and
$$
\vert \mathrm{M3b}_{\mathrm{R}} \rangle =
\vert \mathrm{R},+1\rangle.
$$
Equivalently, in the linearly polarized vector-mode basis, the same M3 subspace can be written as
$$
\vert \mathrm{M3a}_{+} \rangle =
\frac{1}{\sqrt{2}}
\left(
\vert \mathrm{R},+1\rangle
+
\vert \mathrm{L},-1\rangle
\right),
$$
and
$$
\vert \mathrm{M3b}_{-} \rangle =
\frac{1}{\sqrt{2}}
\left(
\vert \mathrm{R},+1\rangle
-
\vert \mathrm{L},-1\rangle
\right).
$$

This mode structure directly determines the possible structured single-photon states generated from quantum-dot transitions. For M2 and M4, left- or right-circularly polarized quantum-dot dipoles can couple to the corresponding circular components of the non-degenerate Bell-like vector modes, generating photons with spin--OAM entanglement. For M3, circularly polarized excitation selectively addresses $\vert \mathrm{M3a}_{\mathrm{R}}\rangle$ or $\vert \mathrm{M3b}_{\mathrm{L}}\rangle$, producing chiral OAM photons with spin--OAM locking. In contrast, linearly polarized excitation can couple to the vector-mode basis $\vert \mathrm{M3a}_{+}\rangle$ or $\vert \mathrm{M3b}_{-}\rangle$, enabling the generation of Bell-like spin--OAM entangled photons. 

This interpretation is consistent with previous studies of polariton micropillars, where the first excited modes were described in a basis of orbital angular momentum and circular polarization states, and the spin--orbit coupling reorganized the degenerate modes into vector modes \cite{real2021chiral}. It is also consistent with the conventional modal analysis of micropillars, where the micropillar cross section can be treated as a finite-radius circular dielectric waveguide and the confined optical eigenmodes are obtained by solving the corresponding electromagnetic boundary-value problem \cite{Reitzenstein2010Quantum}.\\

\noindent \textbf{Fabrication process:}
We use an III-V wafer consisting of a single layer of low-density InAs QDs between 20 (top) and 30 (bottom) GaAs/$\rm Al_{0.9}Ga_{0.1}As$ DBRs grown by molecular beam epitaxy. The deterministically coupled QD-in-micropillars are fabricated by using the hyperspectral fluorescence imaging technique\cite{liu2024super}. First, we fabricate an array of metallic alignment markers on the surface of the sample by using E-beam lithography (EBL), metal evaporation, and standard lift-off processes. After that, a hyperspectral imaging positioning technique is employed to acquire both the spatial positions and emission wavelengths of multiple QDs simultaneously. Then, the positions and diameters of the micropillars are determined based on information obtained by the hyperspectral imaging process. Finally, a large number of deterministically coupled QD-in-micropillar devices are formed by a chlorine-based dry etching. To ensure the spatial matching between the QD and the high-order structured cavity mode, the positions of the pillars are misaligned from the QD by a quarter of the pillar diameter during EBL, deliberately. The devices with emission peaks near-resonant with the p-orbital modes are selected in this work, as shown in Extended Data Fig. E2.\\

\noindent \textbf{Optical measurements:}
The schematic of the setup for optical characterizations is presented in Extended Data Fig. E7. The sample is located in a closed-cycle cryostat with a base temperature of 1.67~K. The cryostat is equipped with a superconducting magnet capable of applying magnetic fields to the QD ranging from -9 T to 9 T. The QDs can be excited by 785 nm continuous wave (CW) or pulsed lasers. Under magnetic field application in the Faraday configuration, the circularly polarized dipole emitters in the QDs undergo splitting and can be tuned across modes M2-M4 by adjusting the magnetic field strength. A customized confocal microscope enables QD spectroscopy and imaging. The quarter-wave plate (QWP), half-wave plate (HWP), and polarizer in the collection path facilitate projection of the emission into the RCP or LCP basis. Emitted photons from the QD are collected via a single-mode fiber and directed to either a spectrometer for spectral analysis or single-photon detectors for correlation measurements. Customized bandpass filters are employed to spectrally select the targeted emissions while suppressing background noise. These filters are grating-based spectral filters, in which the desired wavelength component is coupled into an optical fiber, providing a bandwidth of \(0.061~\mathrm{nm}\) and a transmission efficiency of approximately \(70\%\).

The OAM purity of the emitted single photons was characterized by projecting the emission onto helical phase basis states using vortex wave plates (VWPs). When the QD transition was tuned into resonance with modes M3a and M3b, the emitted photons were sent through VWPs with topological charges $l'$ ranging from $-4$ to $+4$, and the projected single-photon-level spatial images were recorded by an electron-multiplying charge-coupled device (EMCCD).

For a probe order $l'$, the projection coefficient can be written as

$$
c_{l'} =
\iint U(x,y)
\exp\left[i l' \arctan(y/x)\right]
dxdy ,
$$

where $U(x,y)$ is the complex transverse field of the emitted photons. Experimentally, the VWP provides the corresponding helical phase compensation. When the probe order satisfies $l'+l=0$, where $l$ is the incident OAM charge, the vortex phase is cancelled, and a bright on-axis spot appears in the projected far-field image. In contrast, a mismatched projection leaves a residual vortex phase and therefore gives a dark center. The EMCCD directly records the projected cavity-mode profile. The measured central intensity is proportional to $|c_{l'}|^2$ and is used to construct the OAM spectrum. The normalized OAM weight is

$$
P_{l'} =
\frac{|c_{l'}|^2}
{\sum_{l'} |c_{l'}|^2} ,
$$

and the OAM purity is defined as the normalized weight of the dominant OAM component,

$$
\mathcal{P}_{\mathrm{OAM}} =
\frac{|c_{l_{\mathrm{max}}}|^2}
{\sum_{l'} |c_{l'}|^2} .
$$

To verify polarization--OAM entanglement when the QD transition was tuned into resonance with modes M2 and M4, we performed quantum state tomography in the hybrid polarization--OAM basis. The tomography setup consists of a VWP with $|l'|=1$ and a set of polarization-analysis optics, including a quarter-wave plate, a half-wave plate, and a linear polarizer. A complete tomography measurement includes 16 projection settings. The projection intensities were recorded for all settings and used to reconstruct the density matrix of the emitted single-photon state using a maximum-likelihood estimation method. The entanglement fidelity was then obtained by comparing the reconstructed density matrix with the corresponding ideal polarization--OAM entangled state. The measured projection intensities used for the reconstruction are shown in Extended Data Fig. E8.\\

\end{document}